\documentclass[preprint,5p,times,twocolumn]{elsarticle}

\usepackage{graphicx}
\usepackage{epstopdf}
\usepackage{color}
\usepackage{bm}
\usepackage{amssymb}
\usepackage{amsfonts}
\usepackage{amsmath}
\usepackage{float}
\usepackage{siunitx}
\usepackage{multirow}
\newcolumntype{M}[1]{>{\centering\arraybackslash}m{#1}}

\journal{Carbon}
\begin{document}
\begin{frontmatter}

\title{Three-dimensional carbon Archimedean lattices for
    high-performance electromechanical actuators}

\author[a]{Nguyen T. Hung\corref{cor1}} 
\cortext[cor1]{Corresponding author. Tel.:+81 22 795 7754;
Fax: +81 22 795 6447.}
\ead{nguyen@flex.phys.tohoku.ac.jp}
\author[a]{Ahmad R. T. Nugraha}
\author[a]{Riichiro Saito}
\address[a]{Department of Physics, Tohoku University, Sendai 980-8578, Japan}


\begin{abstract}
  We propose three-dimensional carbon (3D-C) structures based on the
  Archimedean lattices (ALs) by combining $sp^2$ bonding in the
  polygon edges and $sp^3$ bonding in the polygon vertices.  By
  first-principles calculations, four types of 3D-C ALs: ($4,8^2$),
  ($3,12^2$), ($6^3$), and ($4^4$) 3D-Cs are predicted to be stable
  both dynamically and mechanically among 11 possible ALs, in which
  the notations ($p_1,p_2,\ldots $) are the indices of the AL
  structures.  Depending on their indices, the 3D-C ALs show distinctive
  electronic properties: the ($4,8^2$) 3D-C is an indirect band-gap
  semiconductor, the ($3,12^2$) 3D-C is semimetal, while the ($6^3$)
  and ($4^4$) 3D-Cs are metals.  Considering the structural
  deformation due to the changes in their electronic energy bands, we
  discuss the electromechanical properties of the 3D-C ALs as a
  function of charge doping.  We find a semiconductor-to-metal and
  semimetallic-to-semiconductor transitions in the ($4,8^2$) and
  ($3, 12^2$) 3D-Cs as a function of charge doping,
  respectively. Moreover, the ($3,12^2$) 3D-C exhibits a $sp^2$-$sp^3$
  phase transformation at high charge doping, which leads to a huge
  30\% irreversible strain, while the reversible strain in the
  ($4,8^2$) 3D-C is up to 9\%, and thus they are quite promising for
  electromechanical actuators.
\end{abstract}

\end{frontmatter}

\section{Introduction}
Scientific and technological attempts to design and synthesize
actuation materials as a building block of artificial muscle have been
quite intensive during the past two decades, with wide potential
applications in biometric machines, robotics, and medical
sciences~\cite{baughman2005playing,madden2006artificial,mirfakhrai2007polymer,hu2010electromechanical}.
Such an actuation material based on carbons was proposed by Baughman
at al. in 1999, who synthesized carbon nanotube (CNT) yarn and
demonstrated that CNT yarn can generate \emph{stress} 100 times that
achievable by natural muscles ($\sim$ 0.35
MPa)~\cite{baughman1999carbon}. Following this initial demonstration,
many studies have sought other carbon actuator materials such as
single, multilayer graphene, graphene oxide, graphitic carbon nitride,
and CNT films via both experimental and theoretical
approaches~\cite{liang2011electromechanical,huang2012application,rogers2011graphene,rogers2011high,wu2015graphitic,li2011superfast,levitsky2004electromechanical}.
However, both CNT yarn and graphene showed a small \emph{strain} of
about 1\% under charge (electron or hole)
doping~\cite{sun2002dimensional,hung2017charge}, while the skeletal
muscles provide a work cycles involving contractions of more than
20\%.  The CNT yarn or the multilayer graphene are also formed by the
weak van der Waals (vdW) forces, which lead to a possible mechanical
failure along the direction of the vdW interactions in these
material~\cite{hung2017charge}. To overcome this problem, we may use
such CNT and graphene structures to form new three-dimensional carbon
(3D-C) materials that still inherit their superb properties while
simultaneously avoiding vdW interaction between different tubes or
layers.

Carbon is one of the most versatile elements due to its ability to
form $sp-$, $sp^2-$, and $sp^3-$hybridized bonds, resulting in various
allotropes that have already been experimentally identified, such as
graphite, diamond, fullerene, CNT, and
graphene~\cite{saito1998physical}. Recently, computational modeling
based on the first-principles density functional theory (DFT) and
molecular dynamics opens the possibility to predict more carbon
allotropes.  Some new 3D-C materials, such as 3D-C
honeycomb~\cite{park2000electronic}, T-carbon~\cite{sheng2011t},
Y-carbon~\cite{jo2012carbon}, L-carbon~\cite{yang2013theoretical}, and
cubane-based 3D porous carbon~\cite{srinivasu2012electronic} have been
either observed or proposed, where all these materials are found to be
dynamically, mechanically, and thermally stable.  Furthermore, the
elastic constants of some of these 3D-C structures were studied.  Zhao
et al.~\cite{zhao2011three} showed that 3D-C structures have large
Young's moduli, large tensile strength, and low density.  Recently,
experimentalists have successfully synthesize a kind of 3D-C
structures in forms of carbon honeycomb by deposition of
vacuum-sublimated graphite~\cite{krainyukova2016carbon} and nanoporous
carbon made by buckybowl-like
nanographene~\cite{nishihara2009possible}.  This fact shows a great
potential in designing the 3D-C materials and structures while
retaining low density and excellent mechanical properties for 
electromechanical actuators. In this sense, a systematic theoretical 
design could be the first step to suggest possible structures 
suitable as electromechanical actuators so that experimentalists 
can synthesize the recommended structures.

\begin{figure*}[t!]
\begin{center}
\includegraphics[width=13cm]{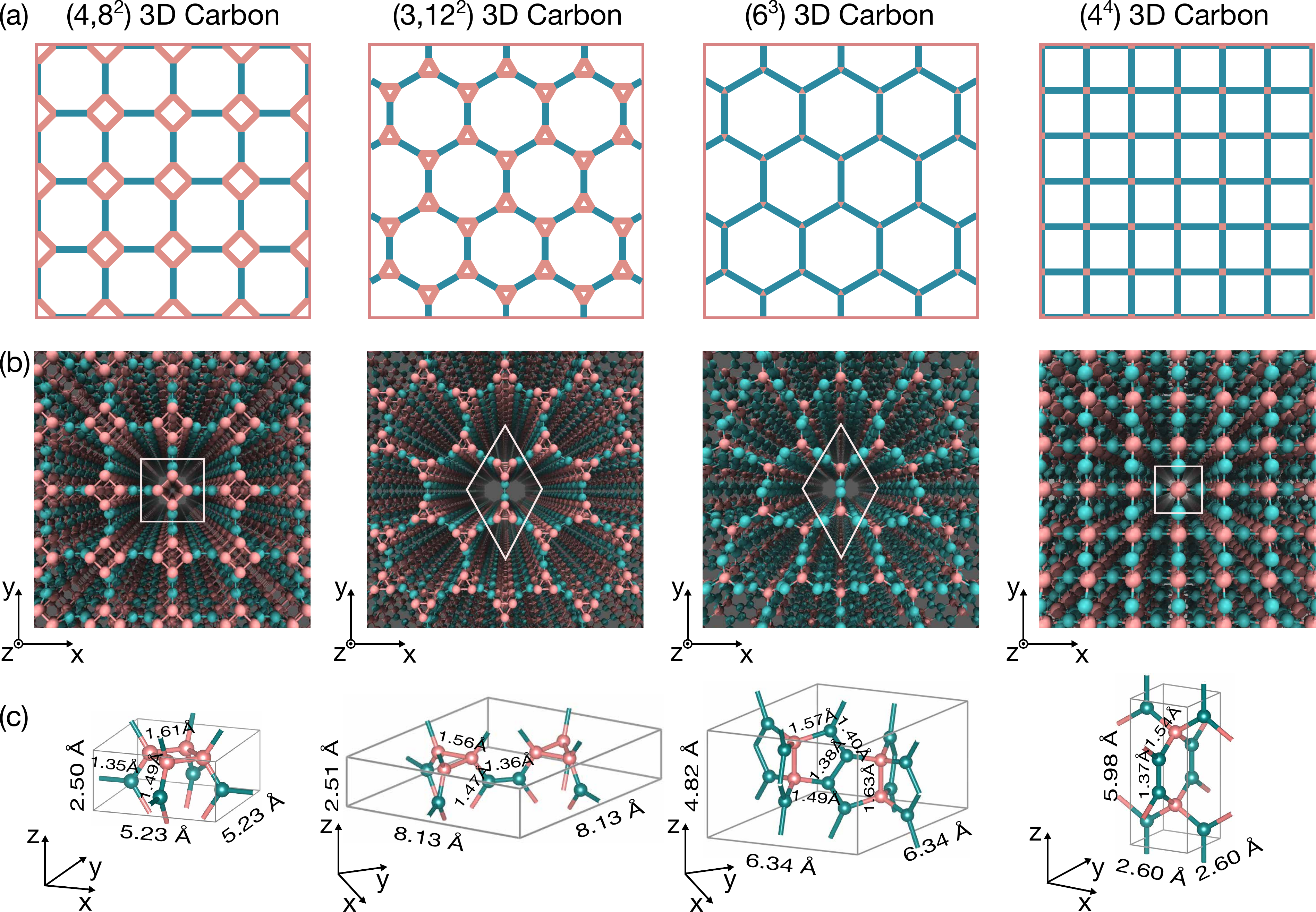}
\caption{(a) The Archimedean lattices. (b) Perspective view of the
  3D-C ALs structures in three-dimensional space. (c) The unit cell of
  the 3D-C ALs showing the lattice constants and the C-C bond lengths.
  Red atoms are $sp^3$-hybridized, while green atoms are
  $sp^2$-hybridized.}
\label{model}
\end{center}
\end{figure*}

Inspired by the structure of the Archimedean lattices (ALs), defined
as a complete set of lattices having all vertices are equivalent, we
design a class of 3D-Cs by a combination of $sp^2$ bonding in the
polygon edges and $sp^3$ bonding in the polygon vertices.  In this
work, considering efficient computational time, 4 of 11 possible 3D-C
ALs are calculated using the first-principles DFT calculations.
Hereafter, the 4 possible 3D-C ALs are referred to as ($4,8^2$),
($3,12^2$), ($6^3$), and ($4^4$) 3D-Cs, in which the notations
($p_1,p_2,\ldots $) are the indices of the AL structures based on the
types of polygons connected at a given vertex (see Fig.~\ref{model}a).
We find that these 3D-C ALs are both dynamically and mechanically
stable because all of their phonon frequencies and mechanical
stability conditions are positive, respectively. We will discuss the
electromechanical properties of the 3D-C ALs as a function of charge
doping, for both electron and hole doping cases, by considering the
structural deformation due to the changes in their electronic energy
bands.  Three important physical phenomena in the doped 3D-C ALs are
found from the calculations: (1) a semiconductor-to-metal and
semimetallic-to-semiconductor transitions in the ($4,8^2$) and
($3,12^2$) 3D-Cs, respectively, (2) a structural transformation
between $sp^2$-$sp^3$ phases under heavy doping in the ($3,12^2$)
3D-C, and (3) a large change of reversible strain up to 5\%, essential
for the artificial muscle applications.

\section{Calculation methods}
We perform first-principles calculations to determine the total energy
and the electronic structure of the 3D-C ALs using Quantum
ESPRESSO~\cite{giannozzi2009quantum}.  The
Rabe-Rappe-Kaxiras-Joannopoulos ultrasoft pseudopotential with an
energy cutoff of 60 Ry is chosen for the expansion of the plane
waves~\cite{rappe1990optimized}.  Note that the choice of 60 Ry energy
cutoff is sufficient for converging the total energy calculation.  The
exchange-correlation energy is evaluated by the general-gradient
approximation using the Perdew-Burke-Ernzerhof (PBE)
function~\cite{pseudo}. In our simulation, the $\mathbf{k}$-point
grids in the Brillouin-zone are employed according to the
Monkhorst-Pack scheme, where $\mathbf{k}$ is the electron wave vector.
$9\times9\times13$, $5\times5\times13$, $7\times7\times9$, and
$16\times16\times6$ $\mathbf{k}$-points are used for ($4,8^2$),
($3,12^2$), ($6^3$), and ($4^4$) 3D-Cs, respectively. To obtain
optimized atomic configurations of 3D-C ALs, the atomic positions and
cell vectors are fully relaxed using the
Broyden-Fretcher-Goldfarb-Shanno minimization
method~\cite{broyden1970convergence,fletcher1970new,goldfarb1970family,shanno1970conditioning}
until all the Hellmann-Feynman forces and all components of the stress
are less than 5e$^{-4}$ Ry/a.u. and 5e$^{-2}$ GPa, respectively, which
are adequate for the present work. To examine the dynamically stable
of the 3D-C ALs, phonon dispersions are computed using the density
functional perturbation theory (DFPT) within the linear response
approximation~\cite{baroni2001phonons}. The dynamical matrices are
calculated on a $4\times4\times6$, $2\times2\times6$,
$3\times3\times4$, and $8\times8\times3$ $\mathbf{q}$-points for
($4,8^2$), ($3,12^2$), ($6^3$), and ($4^4$) 3D-Cs, respectively.

To examine the mechanically stable of the 3D-C ALs, we use the
Thermo-pw code~\cite{dal2016elastic} to calculate the elastic
constants $C_{ij}$. The calculated elastic constants are derived from
the finite difference approach and are by default averaged over the
entire unit cell volumes. From the point of view of elasticity theory,
it is well-recognized that the value of $C_{ij}$ is related to the
equivalent volume of the crystal.  To obtain the equivalent volume of
the 3D-C ALs, the wall thickness, $d$, in the polygon vertices is
considered as the interlayer spacing of graphite and multi-walled
carbon nanotubes in nature based on the van der Waals interactions
($d=3.4$ \AA)~\cite{hung2016intrinsic}, where $d$ is assumed to be
independent of the small strain and charge
doping~\cite{hung2017charge}.  From the calculated $C_{ij}$, the
polycrystalline corresponding bulk modulus $B$ and shear modulus $G$
are calculated using the Voigt-Reuss-Hill
approximation~\cite{anderson1963simplified}, then Young's modulus $E$
and Poisson's ratio $\nu$ are obtained by the following relations,
$E=(9G\cdot B)/(3B+G)$, and $\nu=(3B-2G)/(6B+2G)$, respectively.

To discuss the electromechanical actuation of the 3D-C ALs, the
geometry optimization is performed with the charge doping level,
ranging from $-0.15$ to $0.15$ electron per carbon atom, in which the
electron (hole) doping is simulated by adding (removing) electrons to
the unit cell with the same amount of uniformly positive (negative)
charge in the background so as to keep the charge neutrality.

\begin{table}[t]
  \caption{Number of atoms in unit cell $N$, lattice parameters $a$
    (\AA) and $c$ (\AA), and bond lengths (\AA) of the 3D-C Als.}
\centering  
\begin{tabular}{l c c c M{2.5em} M{2.5em} M{2.5em}}
\hline\hline
Structure & $N$& $a$ & $c$ & $d_{sp^2-sp^2}$  & $d_{sp^2-sp^3}$ &$d_{sp^3-sp^3}$ \\ 
\hline
$(4,8^2)$  & 8 & 5.23& 2.50& 1.35& 1.49 &1.61\\ 
$(3,12^2)$ & 12& 8.13& 2.51& 1.36& 1.47 &1.56\\ 
$(6^3)$    & 16& 6.34& 4.82& 1.38 1.40 &1.49 1.57&1.63\\
$(4^4)$    & 6 & 2.60& 5.98& 1.37& 1.54 & -\\
\hline\hline
\end{tabular}
\label{table:model}    
\end{table}

\section{Results and Discussion}

\subsection{Structural properties}
In Figs.~\ref{model}a-c, we illustrate crystal structures of our
proposed 3D-C ALs.  Each AL is defined by a set of integers
($p_1,p_2,\ldots $) indicating, in a unit cell, the type of polygons
meeting at a given vertex (see Fig.~\ref{model}a).  When a polygon
appears more than one time consecutively, e.g., ($\ldots,p,p,\ldots$),
we abbreviate the notation by writing ($\ldots,p^2,\ldots$). For
example, ($4,8^2$) 3D-C means that each vertex is surrounded
sequentially by one square and two octagons, as shown in
Figs.~\ref{model}a and~\ref{model}b. This property has made them
useful in the study of mathematics~\cite{martinez1973archimedean},
crystallization~\cite{takeda1965compound}, as well as
metamaterials~\cite{jovanovic2008refraction}. Now, the designing
principle of our 3D-C structures is that the polygon vertices are
formed by diamond-like $sp^3$ bond, while the polygon edges consist of
flattened $sp^2$ bond such like that in graphene sheets, as shown in
Fig.~\ref{model}b.  For simplicity, we focus on four allotropes of the
3D-C ALs that are found to be stable theoretically, namely ($4,8^2$),
($3,12^2$), ($6^3$), and ($4^4$) 3D-Cs.  We note that some ALs such as
($3^4,6$) or ($3^2,4,3,4$) might not be suitable to be a 3D-C
structure because the condition for the $sp^3$ bonding of carbon atom
at each vertex is not satisfied~\cite{saito2003carbon}.  The details
of the crystal structures, including the number of atoms in unit cell,
lattice parameters and bond lengths between $sp^2$ (green atoms) and
$sp^3$ bonded atoms (red atoms), are shown in Table~\ref{table:model}
and Fig.~\ref{model}c.

\begin{figure} [t!]
\begin{center}
\includegraphics[width=8cm]{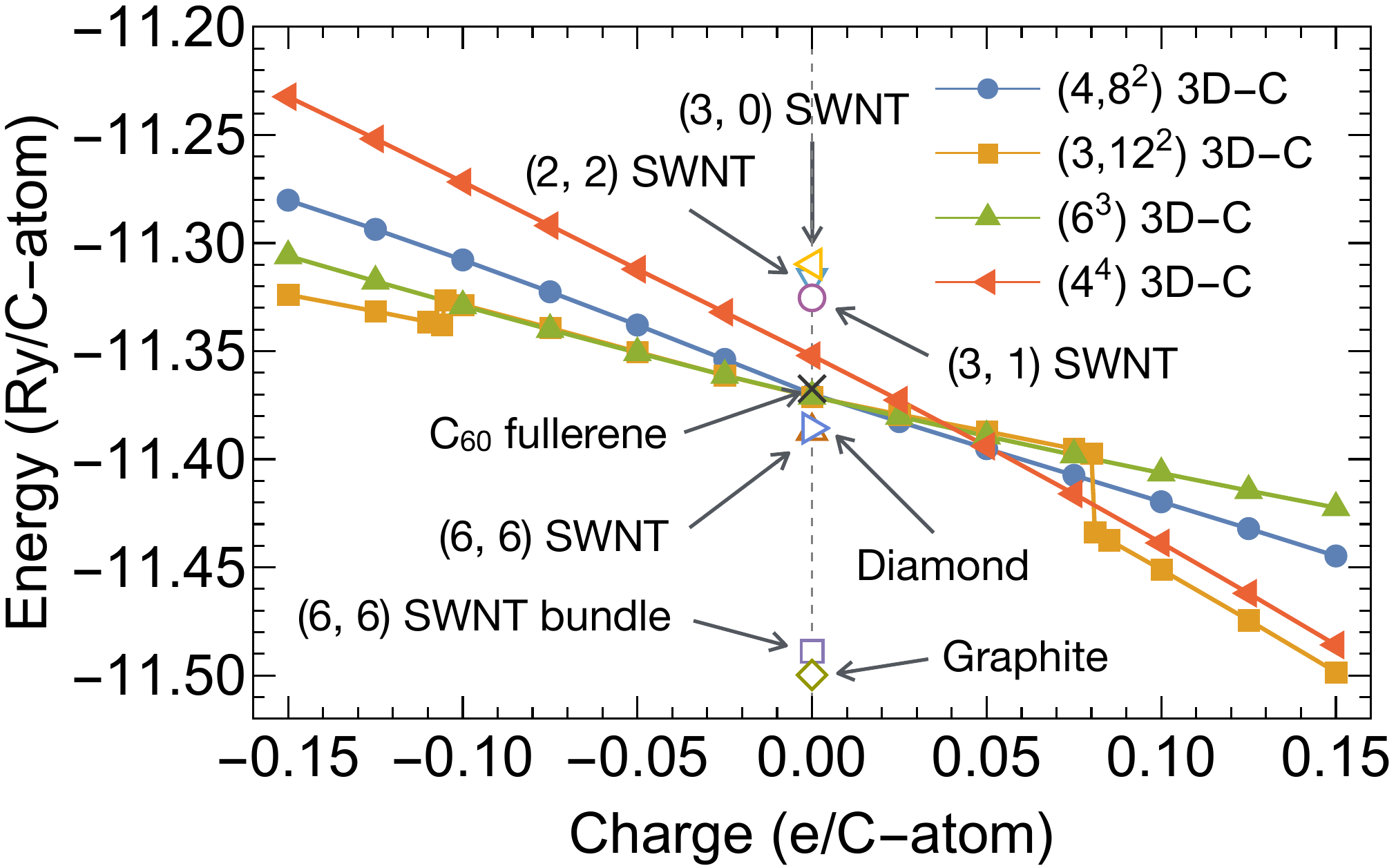}
\caption{Optimized total energy per C-atom of the 3D-C ALs as a
  function of charge (electron and hole) doping per C-atom. At the
  charge $q=0.00$ $e$/C-atom, total energies of individual ($6,6$)
  SWNT~\cite{hung2016intrinsic}, ($6,6$) SWNT
  bundle~\cite{hung2017charge}, smallest carbon nanotubes including
  ($2,2$), ($3,0$) and ($3,1$) SWNTs~\cite{hung2016intrinsic},
  graphite, diamond, C$_{60}$ fullerene are also calculated and plotted here for comparison.}
\label{energy}
\end{center}
\end{figure}

\subsection{Thermodynamics stability}
In Fig.~\ref{energy}, we show the optimized total energy per carbon
atom of the 3D-C ALs as a function of charge doping $q$ ranging from
$-0.15$ to $0.15$ electron per carbon atom.  This charge range is
appropriate for our calculations because the limit of the
experimentally accessible charge is 0.3 $e$/C-atom for
graphite~\cite{sun2002dimensional}. In the neutral condition, the
total energy of the 3D-C ALs is close to that of C$_{60}$ fullerene,
3D diamond, and individual ($6,6$) single wall carbon nanotube
(SWNT)~\cite{hung2016intrinsic}. On the other hand, the 3D-C ALs are
less stable than ($6,6$) SWNT bundle and
graphite~\cite{hung2017charge}, but more stable than the smallest
carbon nanotubes such as ($2,2$), ($3,0$) or ($3,1$)
SWNTs~\cite{hung2016intrinsic}. In fact, experimental evidence has
shown that smallest SWNTs with diameters of about 0.4 nm exist as an
inner core of multiwall carbon nanotubes and can also be found in the
nanosize channels of porous
materials~\cite{qin2000materials,wang2000materials}. Moreover,
Krainyukova et al.~\cite{krainyukova2016carbon} have recently reported
a stable 3D-C by the deposition of vacuum-sublimated graphite that
points out to the possibility of obtaining a ($6^3$) 3D-C structure.
These experimental observations thus support a realization of the
other 3D-C ALs in the near future.  Of the four 3D-C ALs, the ($4^4$)
3D-C is the least stable, while the other 3D-C ALs have relatively
similar energies, especially at the neutral condition.  The total
energy per atom in the 3D-C ALs monotonically decreases with
increasing the charge doping except for the ($3,12^2$) 3D-C.  In the
case of ($3,12^2$) 3D-C, a finite jump of total energy per atom at
electron ($q=-0.106$ $e$/C-atom) doping and hole ($q=0.081$
$e$/C-atom) doping suggests $sp^2$-$sp^3$ phases transformation, which
will be discussed later based on the charge density calculation.

\begin{figure}[t!]
\begin{center}
\includegraphics[width=85mm]{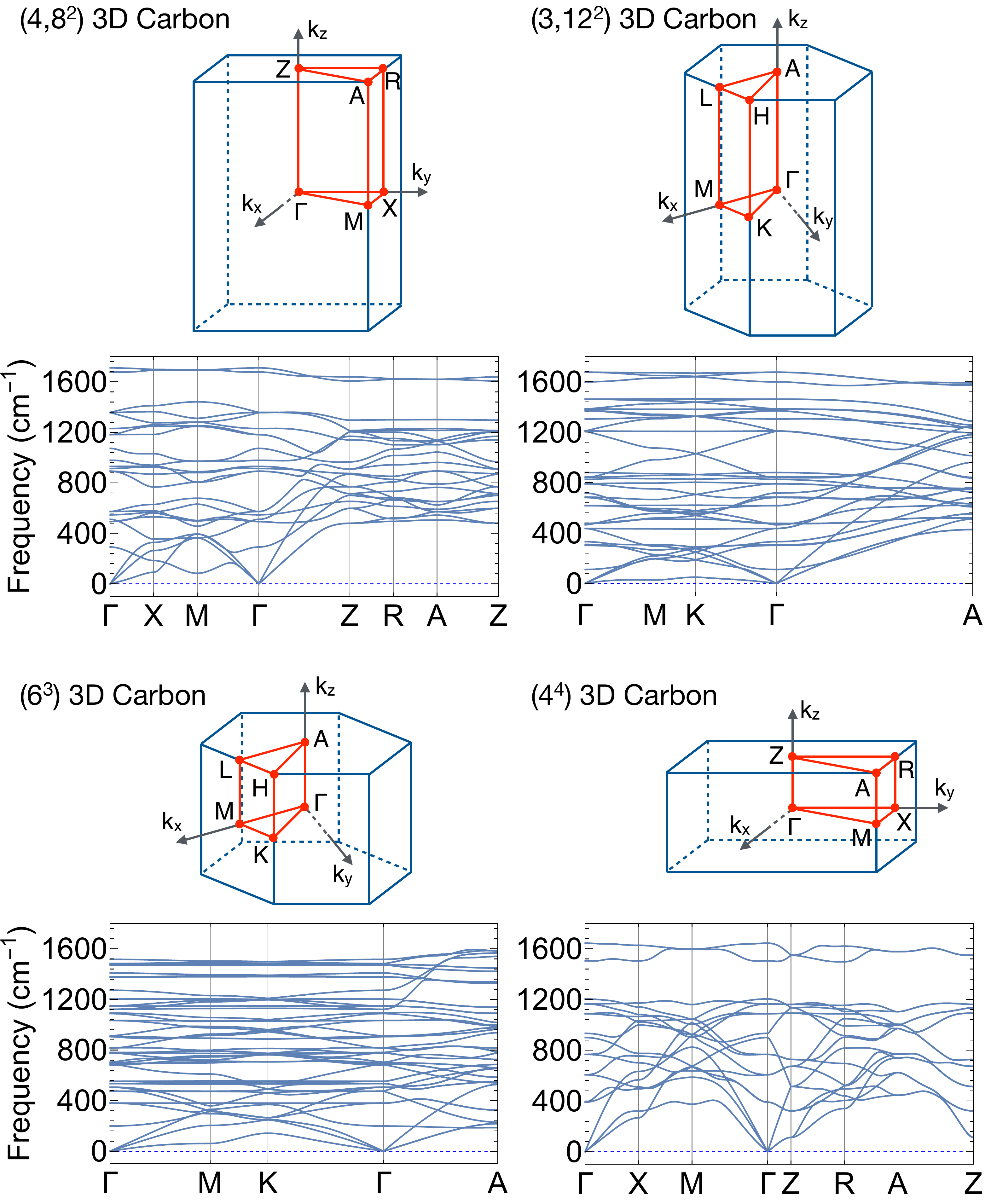}
\caption{Phonon dispersions, high-symmetry points, and lines in the corresponding
Brillouin zone of four different 3D-C ALs.}
\label{phonon}
\end{center}
\end{figure}

\begin{table*}[t]
\caption{Elastic constants $C_{ij}$ (GPa), bulk modulus $B$ (GPa), shear modulus $G$ (GPa), Young's modulus $Y$ (GPa), Poisson's ratio $\nu$, and mass density $\rho$ (g/cm$^3$) of the 3D-C Als, graphite, and diamond.}
\centering  
\begin{tabular}{l c c c c c c c c c c c}
\hline\hline
Structure & $C_{11}$& $C_{33}$& $C_{44}$& $C_{66}$& $C_{12}$& $C_{13}$&$B$& $G$& $Y$& $\nu$& $\rho$\\ 
\hline
$(4,8^2)$  & 651& 1026& 260& 30& 90& 67& 303& 179& 449& 0.25& 2.33\\ 
$(3,12^2)$ & 372& 1004& 261& 45&283& 86& 286& 138& 357& 0.29& 1.67\\  
$(6^3)$    & 346& 1044& 178& 46&254& 56& 263& 124& 320& 0.30& 1.90\\
$(4^4)$    & 704& 1151&  76& 72& 45& 91& 325& 155& 402& 0.30& 2.97\\
grahite$^a$&1060&   36&   4&440& --& 15&   8& 440&1020& 0.16& 2.30\\
diamond$^a$&1070& 1070& 575&575& --&125& 442& 576&1063& 0.10& 3.63\\
\hline\hline
\multicolumn{7}{l}{$^a$ Reference~\cite{zhao2011three,lu1997elastic}.}
\end{tabular}
\label{table:3D-C}    
\end{table*}

\begin{figure*}[t!]
\begin{center}
\includegraphics[width=12cm]{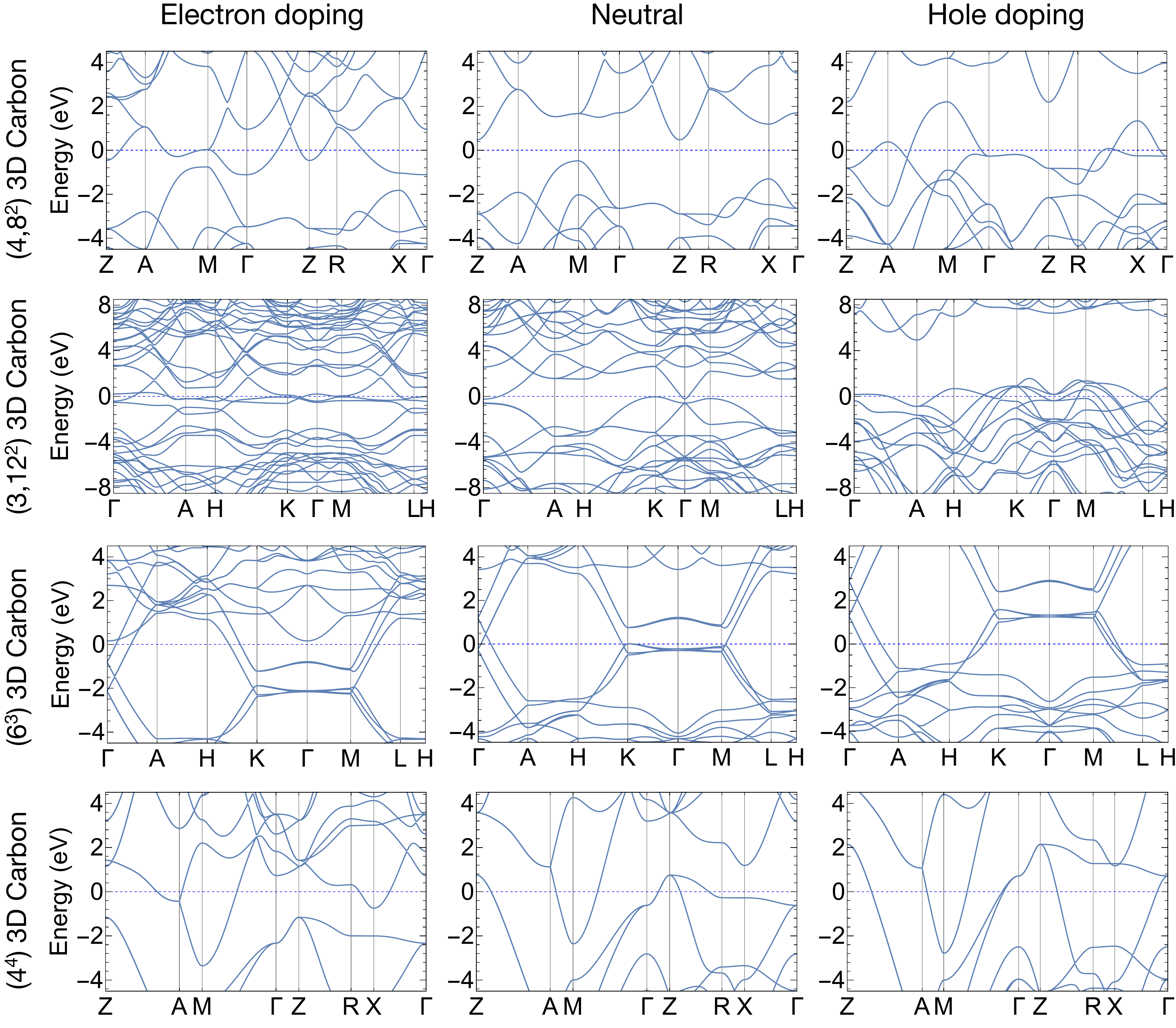}
  \caption{\label{fig:S1} Energy band structures of the 3D-C ALs with
    different electron doping ($q=-0.15$ $e$/C-atom) and hole doping
    ($q=+0.15$ $e$/C-atom) including those with the neutral condition
    ($q=0.00$ $e$/C-atom).  The Fermi energy (dashed
    line) is set to zero for all plots.}
\label{bands}
\end{center}
\end{figure*}

\subsection{Dynamical and mechanical stability}
\label{stability}
In Fig.~\ref{phonon}, we show phonon dispersions along the high-symmetry
directions in the corresponding Brillouin zone of the four 3D-C ALs to 
discuss their lattice dynamics.  All real phonon energies
(positive eigenvalues from the Hessian matrix) confirm 
that the 3D-C ALs are dynamically stable.  Three distinct
acoustic modes including the in-plane longitudinal (LA), in-plane
transverse (TA) and out-of-plane (ZA) modes exhibit linear dispersions
near the $\Gamma$ point, these modes are dominated by $sp^2$ bonding
in the polygon edges.  Note that, the ZA mode in the 3D-C ALs is
unlike that in graphene sheet since the graphene ZA mode is quadratic
dispersion~\cite{si2012electronic}. The highest phonon frequency in
3D-C ALs reaches 1600 cm$^{-1}$, comparable to that in
graphite~\cite{maultzsch2004phonon} and diamond~\cite{warren1967lattice}. 

Next, we investigate the mechanical stability of the 3D-C ALs. To
guarantee the positive strain energy during lattice distortion, the
tensor components of the linear elastic constant ($C_{ij}$) of a
stable crystal have to obey the Born
criteria~\cite{born1954dynamical}. In the ($4,8^2$) and ($4^4$) 3D-Cs,
there are six independent elastic constants due to tetragonal
symmetry, while in the ($3,12^2$) and ($6^3$) 3D-Cs there are only
five constants due to hexagonal symmetry, in which
$C_{66}=(C_{11}-C_{12})/2$.  According to the Born criteria, the
elastic constants of the tetragonal and hexagonal crystals have to
satisfy the following conditions~\cite{mouhat2014necessary}:
$C_{11} > |C_{12}|$, $2C_{13}^2 < C_{33}(C_{11}+C_{12})$,
$C_{44} > 0$, and $C_{66} > 0$ (the condition for $C_{66}$ is
redundant in the hexagonal case).  The elastic constants can be
derived by fitting the strain-stress curves associated with uniaxial
and equibiaxial strains, as implemented in Thermo-pw
code~\cite{dal2016elastic}. The calculated $C_{ij}$ values of the 3D-C
ALs are listed in Table~\ref{table:3D-C} at the neutral charge.  These
constants obey all of the criteria above, indicating that all the
four 3D-C ALs are mechanically stable.

\begin{figure}[t!]
\begin{center}
\includegraphics[width=8cm]{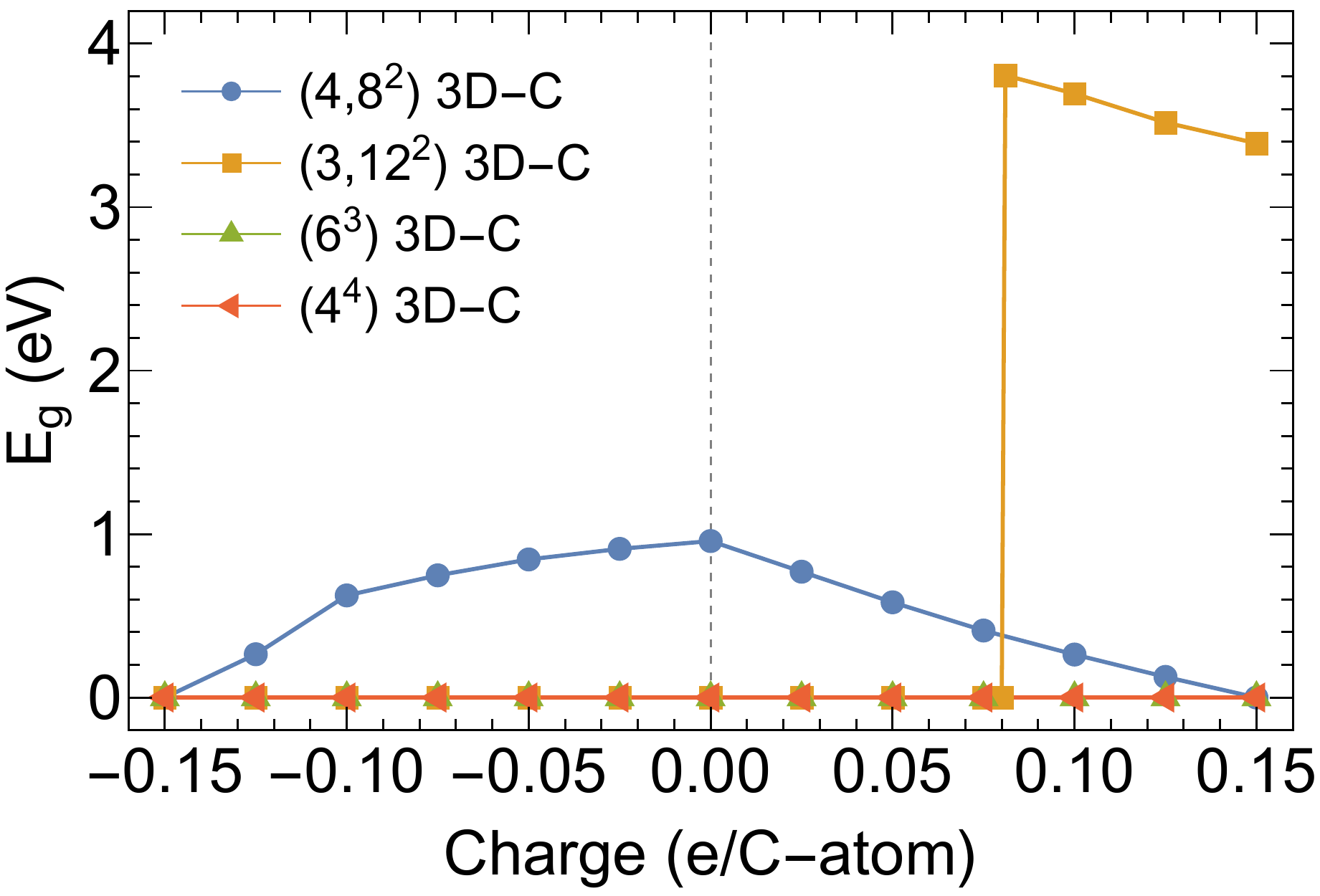}
\caption{Energy band gap $E_g$ plotted as function
  of charge doping of four different 3D-C ALs.
\label{band-gap}}
\end{center}
\end{figure}

\begin{figure}[t!]
\begin{center}
\includegraphics[width=8cm]{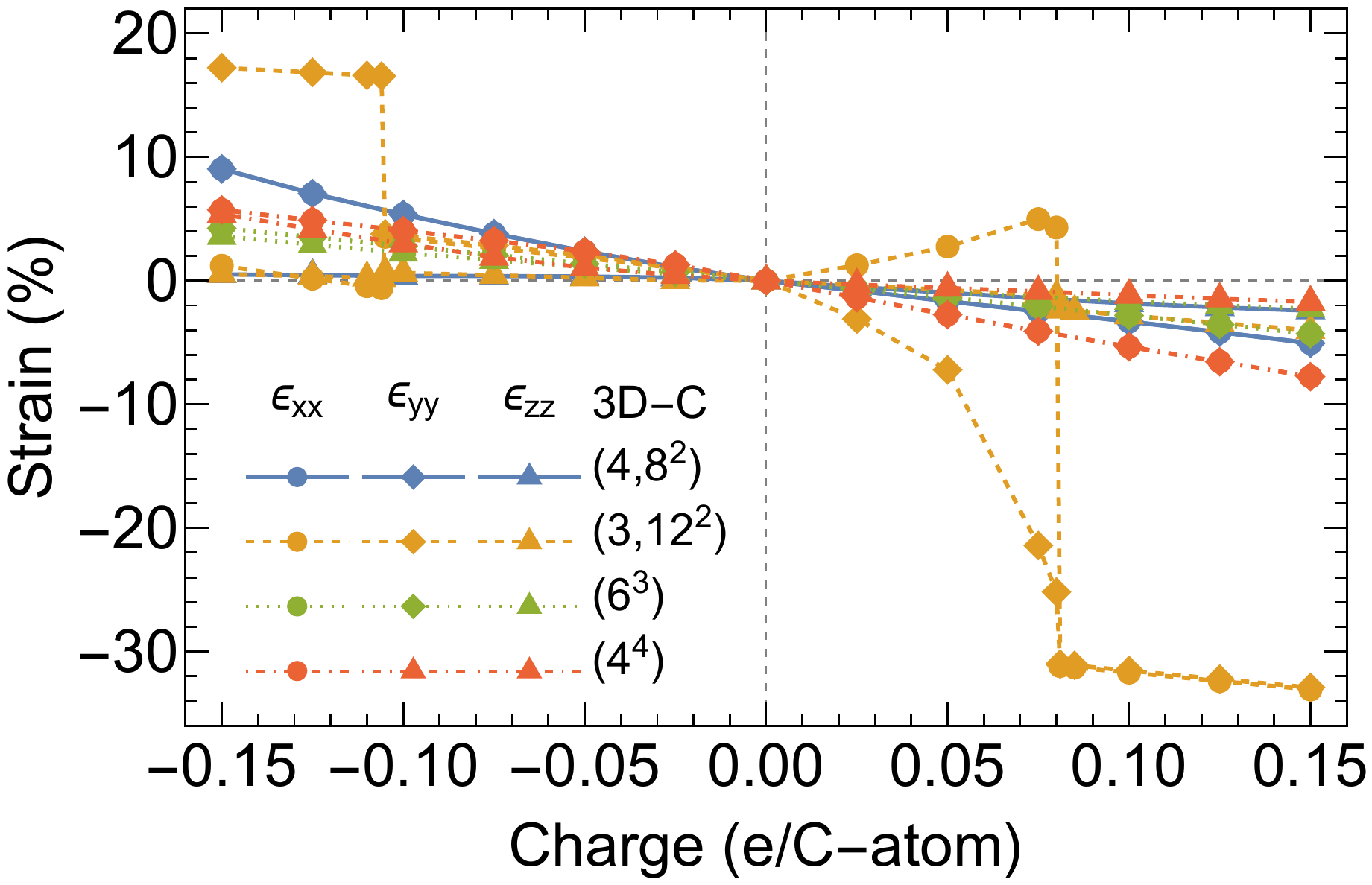}
\caption{Strain  as function of charge doping of four different 3D-C ALs.
  \label{strain}}
\end{center}
\end{figure}

\begin{figure}[t!]
\begin{center}
\includegraphics[width=7cm]{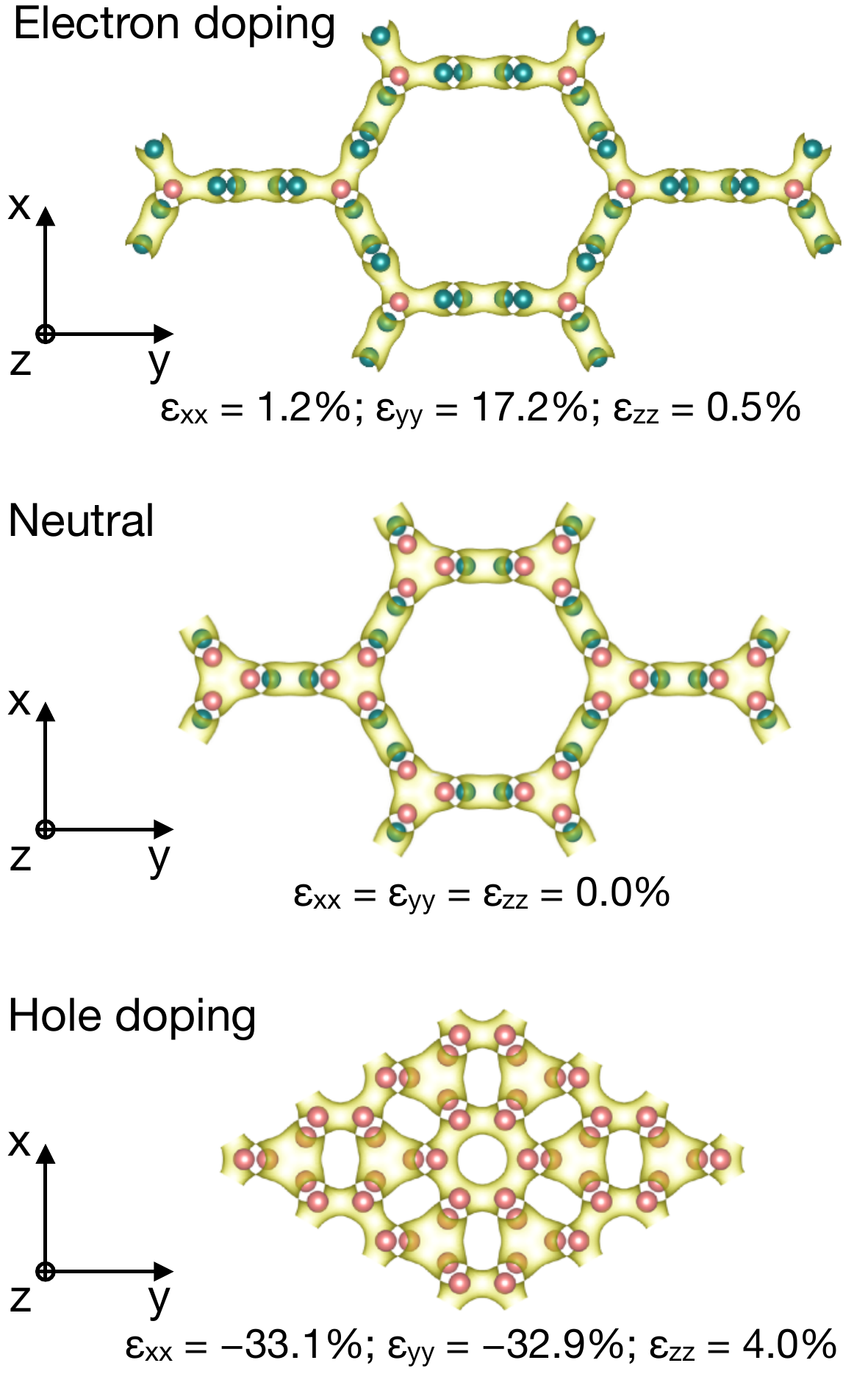}
\caption{Charge density
  isosurface (0.13 $e$/a.u.$^3$) of the ($3,12^2$) 3D-C at neutral,
  heavy electron and hole dopings cases.  Red atoms are $sp^3$
  hybridized and blue atoms are $sp^2$ hybridized.  At heavy electron
  and heavy hole doping regimes, the ($3,12^2$) 3D-C is transformed to
  ($6^3$) 3D-C and ($3, 4, 6, 4$) 3D-C, respectively.
\label{charge}}
\end{center}
\end{figure}

\subsection{Electronic properties}
In Fig.~\ref{bands}, we show the calculated electronic structures of
the 3D-C ALs along the high-symmetry directions in the corresponding
Brillouin zone (see Fig.~\ref{phonon}) for neutral and charge doping
states.  For the neutral case, we find that the ($4,8^2$) 3D-C is an
indirect-gap semiconductor with a band gap of about 0.96 eV, the
($3,12^2$) 3D-C is a semimetal, while the ($6^3$) and ($4^4$) 3D-Cs
are metals. The unusual electronic properties of the 3D-C ALs might
originate from $sp^2$ and $sp^3$ bonded atoms (green and red atoms
in Fig.~\ref{model}) at the Fermi level.  For the ($6^3$) and ($4^4$)
3D-Cs, the number of $sp^3$ bonded atoms is smaller than that of
$sp^2$ bonded atoms in the unit cell. Therefore, most of the
conducting electrons at the Fermi level are coming from the 2$p_z$
orbitals of $sp^2$ bonded carbon atoms like that in graphene, which lead to the metallic properties of the ($6^3$) and ($4^4$) 3D-Cs.  As for the ($4,8^2$) and
($3,12^2$) 3D-Cs, the numbers of $sp^2$ and $sp^3$ bonded atoms in the
unit cell are the same. Therefore, both $sp^2$ and $sp^3$ bonded atoms
contribute to the electrons at the Fermi level, which result in
semiconducting ($4,8^2$) and semimetallic ($3,12^2$) 3D-Cs.  Through a
detailed analysis of the energy band structures under the charge
doping, we can see that the energy bands change significantly under
both heavy electron and hole dopings, which could affect the
mechanical properties of the 3D-C ALs.

In Fig.~\ref{band-gap}, we show the energy band gap $E_g$ as a
function of charge doping $q$.  For the semiconducting ($4,8^2$) 3D-C,
$E_g$ decreases with both electron and hole doping cases, leading to a
semiconductor-to-metal transition at $q=\pm0.15$ $e$/C-atom.  For the
semimetallic ($3, 12^2$) 3D-C, a semimetallic-to-semiconductor
transition occurs when $q$ is adjusted up to $0.081$ $e$/C-atom, while
the ($6^3$) and ($4^4$) 3D-Cs retain their metallic states under
charge doping. It is thus clear that the rigid band model, in which
the effective mass and the band gap are fixed while Fermi level is
shifted by charge doping, is not valid for the presented cases. As
shown in Fig.~\ref{band-gap}, the rigid band model is remarkably more
pronounced for the ($3, 12^2$) 3D-C in both electron and hole dopings
cases and for the ($4,8^2$) 3D-C in the electron doping case. For the
($3, 12^2$) 3D-C, $sp^3$ bonded atoms break into $sp^2$ bonded atoms under heavy electron doping, while $sp^2$ bonded atoms make new $sp^3$ bonded atoms under heavy hole doping (see the charge density in Fig.~\ref{charge}).  The energy band structure thus
changes significantly, which leads to the metallic and semiconducting
properties of the ($3, 12^2$) 3D-C under heavy electron and hole
dopings, respectively.  For the ($4,8^2$) 3D-C, its electromechanical
response shows a higher strain value than the ($6^3$) and ($4^4$)
3D-Cs under heavy electron doping (see Sec.~\ref{sub:strain}), which also
leads to a significant change of the energy band structure.
Therefore, the rigid band model is no longer suitable for the
($4,8^2$) 3D-C under heavy electron doping.


\begin{figure*} [t!]
\begin{center}
\includegraphics[width=13cm]{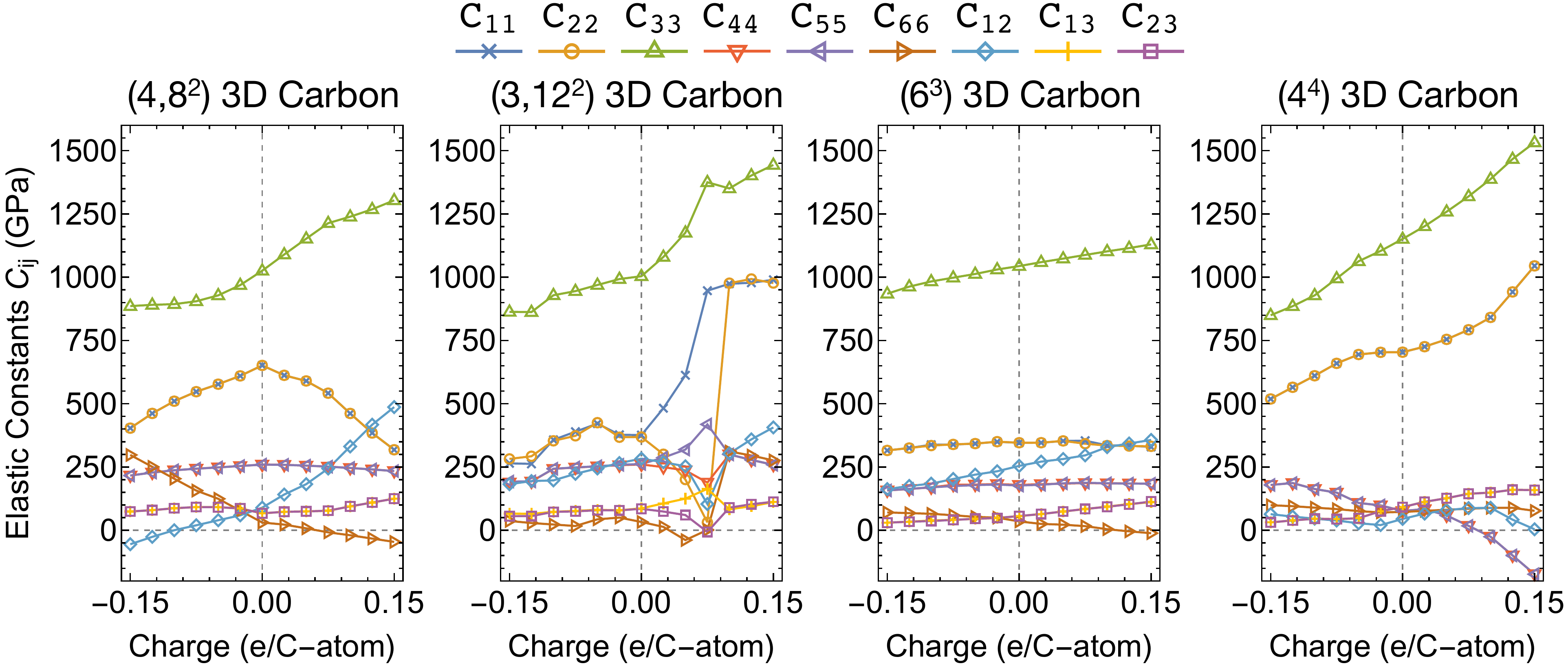}
\caption{Elastic constants $C_{ij}$ of four different 3D-C ALs plotted
  as function of of charge (electron and hole) doping per carbon
  atom.}
\label{elastic}
\end{center}
\end{figure*}

\begin{figure} [t!]
\begin{center}
\includegraphics[width=85mm]{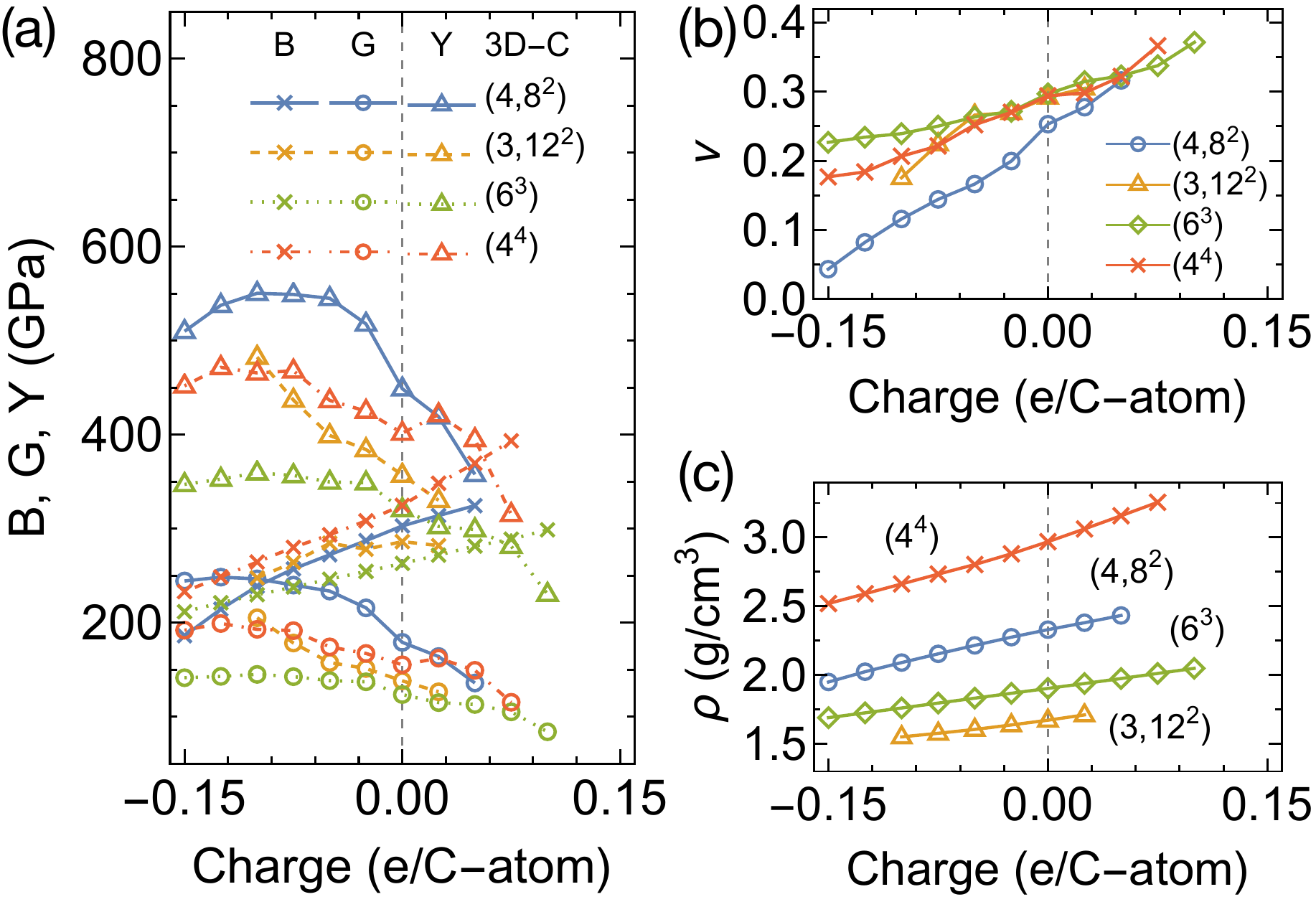}
\caption{(a) Mechanical moduli, including the bulk modulus $B$, the
  shear modulus $G$, and Young's modulus $Y$, (b) Poisson's ratio
  $\nu$, and (c) mass densities $\rho$ of the 3D-C ALs plotted as
  function of charge doping.  Note that we omit some data points of
  heavy doping due to the unstability of certain 3D-C ALs at that
  regime.}
\label{modulus}
\end{center}
\end{figure}

\subsection{Electromechanical properties}
\label{sub:strain}
In Fig.~\ref{strain}, we show the charge-strain relationship of the
3D-C ALs.  The strains of three 3D-C ALs, except the ($3, 12^2$) 3D-C,
are approximately a linear function of charge doping and are isotropic
in $x-y$ plane.  For the ($4,8^2$) 3D-C,
$\varepsilon_{xx}=\varepsilon_{yy}$ is up to 9.0\% ($-5.1$\%) under
the electron (hole) doping at $q=-0.15$ ($0.15$) $e$/C-atom, while
$\varepsilon_{zz}=0.5$\% ($2.4$\%).  For the ($6^3$) 3D-C,
$\varepsilon_{xx}=\varepsilon_{yy}=4.2$\% ($-4.3$\%) and
$\varepsilon_{zz}=3.6$\% ($-2.3$\%) at $q=-0.15$ ($0.15$) $e$/C-atom.
For the ($4^4$) 3D-C, $\varepsilon_{xx}=\varepsilon_{yy}=5.7$\%
($-7.8$\%) and $\varepsilon_{zz}=5.4$\% ($-1.7$\%) at $q=-0.15$
($0.15$) $e$/C-atom.  As for the ($3, 12^2$) 3D-C, the calculation
results give an anisotropic strain under hole doping at $q=0.08$
$e$/C-atom with $\varepsilon_{xx}$ and $\varepsilon_{yy}$ are up to
4.3\% and -25.2\%, respectively.  For further hole doping,
$\varepsilon_{xx}$ and $\varepsilon_{yy}$ show a finite jump that
leads to $\varepsilon_{xx}=\varepsilon_{yy}\sim -31$\% at $q=0.081$
$e$/C-atom.  Similarly, for electron doping, $\varepsilon_{xx}$ and
$\varepsilon_{yy}$ have a finite jump at $q=-1.06$ $e$/C-atom
($\varepsilon_{xx}=-0.6$\% and $\varepsilon_{yy}=16.5$\%).  We argue
that a very high strain of the ($3,12^2$) 3D-C under charge doping is
a result of its chemical bond transformation.  Based on the charge
density plot in Fig.~\ref{charge}, $sp^3$ bonds change to $sp^2$ bonds
under heavy electron doping, while $sp^2$ bonds change to $sp^3$ under
heavy hole doping in the ($3,12^2$) 3D-C.  However, these $sp^2$-$sp^3$
transformations are not reversible.  At heavy electron and heavy hole
doping regimes, the ($3,12$) 3D-C is transformed to ($6^3$) 3D-C and
($3, 4, 6, 4$) 3D-C, respectively.  Nevertheless, the ($3,12^2$) 3D-C
is a stable structure without charge doping.

To understand the stability of the 3D-C ALs under charge doping, the
mechanical stability is investigated by the Born
criteria~\cite{born1954dynamical} at different charge states.
According to the Born criteria, the elastic constants $C_{ij}$ of a
stable 3D-C AL structure has to satisfy the conditions
$C_{11} > |C_{12}|$, $2C_{13}^2 < C_{33}(C_{11}+C_{12})$,
$C_{44} > 0$, and $C_{66} > 0$ (see Sec.~\ref{stability}). In
Fig.~\ref{elastic}, we calculate $C_{ij}$ as a function of charge
doping.  We find that the ($4,8^2$), ($6^3$) and ($4^4$) 3D-Cs are
unstable under heavy hole doping because both $C_{44}$ and $C_{66}$
are less than zero, resulting in a negative strain energy.  On the
other hand, the ($3,12^2$) 3D-C is stable under both heavy electron
and hole doping cases but its structure shows an irreversible
$sp^2$-$sp^3$ transformations under heavy charge doping (see
Fig.~\ref{charge}).  Having confirmed the stability of the 3D-C ALs,
we then systematically study their mechanical properties.

Fig.~\ref{modulus}a shows the bulk modulus $B$, the shear modulus
$G$ and Young's modulus $Y$ of the \emph{stable} 3D-C ALs as a
function of charge doping.  In Fig.~\ref{modulus}, we omit some data
points of heavy doping due to the unstability of certain 3D-C ALs,
i.e. they become distorted, at that regime.  In the neutral condition,
the ($4,8^2$) 3D-C has the highest $G=179$ GPa and $Y=449$ GPa, while
the ($4^4$) 3D-C has the highest $B=325$ GPa. Both $G$ and $Y$ are
decreased, while $B$ is increased by charge doping.  In the case of a
SWNT~\cite{hung2016intrinsic,wu2008determination}, the axial $Y$ is
about 1 TPa, which is consistent with the in-plane $Y$ of a graphene
layer.  However, the 3D structures of CNT composites, such as bundles,
yarns, and fibers appear to be more flexible and elastic in their
radial directions due to the weak van der Waals interactions between
different tubes.  The experimentally obtained $Y$ values of CNT
composites are only in the range of 0.35-80
GPa~\cite{mirfakhrai2007electrochemical,dalton2003super,coleman2006small}.
Compared with CNTs composites, the $Y$ values of the 3D-C ALs are much
higher by two order-of-magnitude.  Therefore, our designed 3D-C ALs
retain not only the superior mechanical performance in the axial
directions of the tubes, but also give enhanced reinforcement in their
radial directions through the $sp^3$ bond connections at the vertices
of the 3D-C ALs.  The large mechanical moduli of the 3D-C ALs are
important for the artificial muscle application since it could
generate large force per unit area.

Figure~\ref{modulus}b shows Poisson's ratio $\nu$ of the \emph{stable}
3D-C ALs as a function of charge doping.  It is known that Poisson's
ratio is a measure of the Poisson effect, a phenomenon in which a
material tends to expand in the directions perpendicular to the
direction of compression.  In the neutral condition, the $\nu$ values
of three 3D-C ALs are about 0.29, except for the ($4,8^2$) 3D-C which
gives $\nu=0.25$.  These values are larger than those of SWNT
($\nu = 0.07-0.15$) and graphene ($\nu = 0.186$), which is suitable
for muscle.  In the 3D-C ALs, the Poisson effect is also further
enhanced by the charge doping.

Finally, Fig.~\ref{modulus}c shows the mass density $\rho$ of the
\emph{stable} 3D-C ALs as a function of charge doping.  At the neutral
charge, ($4^4$) 3D-C has the highest $\rho=2.97$ g/cm$^3$ and
($3,12^2$) 3D-C has the lowest $\rho=1.67$ g/cm$^3$.  These values are
smaller than that of diamond ($3.633$ g/cm$^3$) and SWNT ($\sim 3.0$
g/cm$^3$), and are close to that of graphite ($2.295$
g/cm$^3$)~\cite{zhao2011three}.  The low density of the 3D-C ALs is
due to their porous structures along the axial $z$ directions.
Although the mass density of 3D-C ALs is higher than that of mammalian
skeletal muscle ($\sim 1$g/cm$^3$), we expect that the designed 3D-C
ALs with larger pore diameters along the $z$ direction will give
ultralow density.  Pang et al.,~\cite{pang2017bottom}, for example,
showed that the density of the ($6^3$) 3D-C can go down to 0.31
g/cm$^3$ with the pore diameter of about 29.7 \AA.  However, both of
the strength and failure strain decrease in ($6^3$) 3D-C with growing
pore diameter~\cite{pang2017bottom}.  We also note that for the 3D-C
ALs with large pore diameter, the number of flattened $sp^2$ bonds at
the polygon edges is much higher than that of $sp^3$ bonds at the
polygon vertices. Therefore, the 3D-C ALs might not be able to
maintain their electromechanical properties when their pore diameters
become larger.

\section{Conclusions}
By first principles DFT calculations, we have shown the feasibility of
constructing stable 3D carbon materials based on the Archimedean
lattices with excellent electromechanical properties. The reversible
strain can be up to 9\% with stable ($4, 8^2$) 3D-C structure 
under heavy electron doping,
which is nine times larger than that of carbon nanotube or graphene. A
very high irreversible strain of about 30\% in the ($3, 12^2$) 3D-C is
also found as a result of $sp^2$-$sp^3$ bond transformation.  Large
mechanical moduli (including the bulk modulus, the shear modulus, and
Young's modulus), large Poisson's effect), and low mass density under
charge doping may be utilized to obtain high-performance actuators
with applications ranging from artificial muscle to biomedical
engineering.  Depending on the index of Archimedean lattices, the 3D-C
ALs also show distinctive electronic properties: the ($4,8^2$) 3D-C is
an indirect band-gap semiconductor, the ($3,12^2$) 3D-C is semimetal,
while the ($6^3$) and ($4^4$) 3D-Cs are metals.  Furthermore, we
expect that the semiconductor-to-metal and
semimetallic-to-semiconductor transitions found in the semiconducting
($4,8^2$) and semimetallic ($3, 12^2$) 3D-Cs, respectively, under
heavy charge doping will open further explorations of the 3D-C ALs in
the near future.

\section*{Acknowledgments}
N.T.H. and A.R.T.N acknowledge the Interdepartmental Doctoral Degree
Program for Multidimensional Materials Science Leaders under the
Leading Graduate School Program in Tohoku
University. R.S. acknowledges JSPS KAKENHI Grant Numbers JP25107005
and JP25286005.

\section*{Supplementary information}
Atomic coordinates (CIF files) of the optimized 3D-C ALs structures
studied in this work.

\section*{References}
\bibliographystyle{elsarticle-num}

\end{document}